\documentclass[aps,twocolumn,preprintnumbers]{revtex4-2}

\usepackage{graphicx}  
\usepackage[caption=false]{subfig}
\usepackage{multirow}
\usepackage{fancyhdr}
\usepackage{longtable}
\usepackage{parskip}
\usepackage[T1]{fontenc}
\usepackage{dcolumn}   

\usepackage{bm}        
\usepackage{amsfonts}  
\usepackage{amsmath}   
\usepackage{amssymb}   
\usepackage{hyperref}

\hypersetup{
    colorlinks=true,
    linkcolor=cyan,
    citecolor=cyan,
    urlcolor=cyan}

\usepackage{float}

\newcommand{\pwisein}{\left\{ \begin{array}{ll}}
\newcommand{\pwiseout}{\end{array}\right.}

\newcounter{defcounter}
\usepackage{amsthm}

\begin{document}
\title[]{Leaky wave characterisation using spectral methods}

\author{Evripides Georgiades}
\email{evripides.georgiades18@imperial.ac.uk}
\author{Michael J.S. Lowe}

\author{Richard V. Craster}
\altaffiliation{Also at: Department of Mathematics, Imperial College London, London, SW7 1AY, United Kingdom}

\affiliation{Department of Mechanical Engineering, Imperial College London, London, SW7 1AY, United Kingdom}

\date{\today} 

\begin{abstract}

Leaky waves are an important class of waves, particularly for guiding waves along structures embedded within another medium; a mismatch in wavespeeds often leads to leakage of energy from the waveguide, or interface, into the medium, which consequently attenuates the guided wave. The accurate and efficient identification of theoretical solutions for leaky waves is a key requirement for the choices of modes and frequencies required for non-destructive evaluation inspection techniques. We choose a typical situation to study:  an elastic waveguide with a fluid on either side. Historically, leaky waves are identified via root-finding methods that have issues with conditioning, or, numerical methods` that struggle with the exponential growth of solutions at infinity. By building upon a spectral collocation method, we show how it can be adjusted to find exponentially growing solutions, i.e. leaky waves, leading to an accurate, fast and efficient identification of their dispersion properties. The key concept required is a mapping, in the fluid region, that allows for exponential growth of the physical solution at infinity, whilst the mapped numerical setting decays. We illustrate this by studying leaky Lamb waves in an elastic waveguide immersed between two different fluids and verify this using the commercially available software Disperse.  

\end{abstract}

\maketitle

\section{Introduction}
Ultrasonic guided waves are of enduring interest for Non-Destructive Evaluation (NDE), where guided waves, such as Lamb waves \citep{Lamb1916OnPlate}, are utilised for the inspection or structural health monitoring of single or multilayered systems of plates or pipes \citep{Alleyne1992OptimizationTechniques,Alleyne2001RapidWaves,Fromme2006OnMonitoring,Long2003AttenuationPipes}; the complexities and characteristics of Lamb wave propagation for different systems and media is of paramount importance in identifying the appropriate choices in terms of frequencies and modes\citep{Wilcox2001ModeInspection}. The route to effective inspection techniques depends critically upon  predictive models focused on identifying the dispersion curves and associated stress and displacement fields which have consequentially been the focus of much attention. 

Leaky waves arise in immersed guiding structures: 
it is common for a structure of interest to consist of an elastic waveguide partially in contact or fully immersed in fluids. In this scenario the Lamb waves change their character as energy is no longer necessarily confined within the boundaries of a solid waveguide.  When coupled to inviscid fluid(s) of infinite extent, two types of guided waves arise \citep{Treyssede2014FiniteWaveguides}. The first are guided waves remaining confined to the solid and immediately-adjacent fluid, because their wave speeds are below the speed of bulk waves in the adjacent fluid; these include, at high frequencies, Scholte waves \citep{Scholte1947TheWaves}; these are not the focus of this article. The second type of guided wave has energy leakage, in the form of acoustic energy radiation from the waveguide into the adjacent fluid(s). This leads to attenuation, i.e. decay in amplitude, along the waveguide, and modes in the fluid(s) whose amplitudes grow exponentially in the direction transverse to the waveguide. These are known as leaky Lamb waves \citep{Chimenti1985LeakyLaminates,Nagy1987ADetection,Nayfeh1988PropagationComposite} and are the focus of this study.

A key role in multiple areas of wave physics is played by leaky waves, for instance in guiding structures in electromagnetism \citep{Snyder1983OpticalTheory} where a mismatch of refractive indices between waveguide and exterior medium leads to energy leakage and analogous leaky waves. A comprehensive recent review \citep{Hu2009UnderstandingRevisited},  in the  electromagnetic waveguiding setting, usefully rationalises the phenomenon of the exponential growth of leaky waves and outlines applications in optics and electromagnetism. In acoustics and elasticity, leaky Rayleigh waves feature prominently in acoustic microscopy \citep{Every1998SurfaceMicroscopy}, in material characterisation of the waveguide \citep{Lefeuvre1998LeakyCharacterization} or of the exterior fluid \citep{Cegla2006FluidSensor}, and  underwater structural acoustics
\citep{Stuart1976AcousticPoles}, while leaky Lamb waves \citep{Chimenti1985LeakyLaminates,Nayfeh1997ExcessLoading} feature in the NDE of composites; there is therefore naturally considerable interest in their accurate identification. 

\begin{figure*}
    \centering
    \includegraphics[scale=0.2]{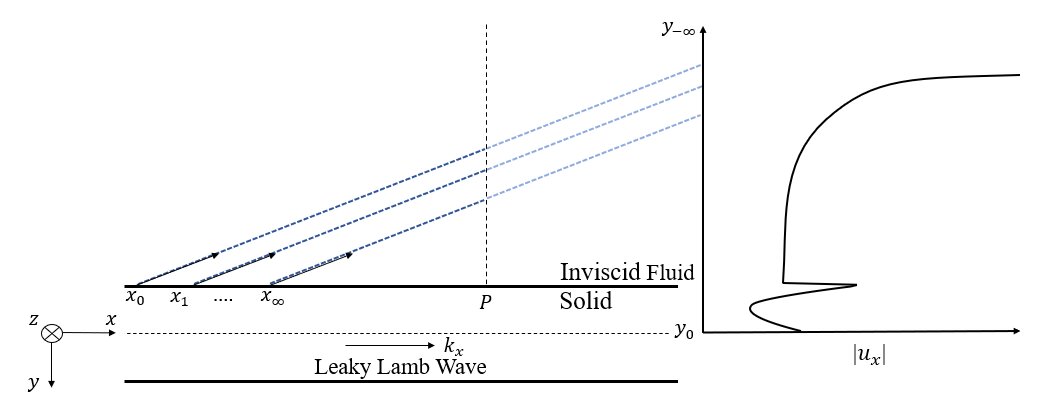}
    \caption{The exponential growth of leaky Lamb waves, as a result of energy radiation at a fluid-solid interface.}
    \label{Leaky wave}
\end{figure*}

An, initially counter-intuitive,  exponential growth of amplitude
at infinity, and corresponding complex-valued roots of dispersion
relations, characterise leaky waves and this results in numerical
issues when cleanly attempting to identify their dispersion characteristics, 
while doing so remains a challenge \citep{Kiefer2019CalculatingInteraction}. 
For those unfamiliar with leaky
waves, the exponential growth of the mode might appear to be perplexing and to contradict energy conservation; a schematic showing the origin of this growth can be seen in Fig. \ref{Leaky wave}. We choose an observation point, $P$, along a waveguide with attentuating waves propagating along the waveguide with propagation constant $k_x$, and then consider the far-field perpendicular to the guide. In the far-field
we are observing the result of energy that leaked from points preceding our observation point: the further we go back along the guide then
the larger the amplitude of the energy that we observe. On this basis
the exponential growth we observe is actually a consequence of energy
conservation.

For non-leaky guided waves, tools for the retrieval of dispersion characteristics are well-developed. Many numerical methods have been employed thus far, with one of the most established ones being Partial Wave Root Finding (PWRF). The general principle of PWRF routines is the production of a matrix formulation of the system in question, through a partial wave description of the problem, and the subsequent employment of root-finding routines for the evaluation of the roots of its characteristic equation and dispersion curves. There exist numerous such matrix formulations for single or multilayered systems, with the Transfer Matrix \citep{Thomson1950TransmissionMedium,Haskell1964TheMedia} and Global Matrix \citep{Knopoff1964AProblems} methods being particularly appealing due to their relatively simple formulations. In fact, the commercially available software DISPERSE \citep{Pavlakovic1997Disperse:Curves} utilises some of these PWRF techniques for the computation of dispersion characteristics of free and immersed systems, providing an excellent verification tool for the results of our study. An overview of the main matrix formulations used in PWRF routines for multilayered waveguides are in Lowe \citep{Lowe1995MatrixMedia}. 

Alas, PWRF routines have their limitations. To start with, their dependence on matrix formulations can result in spurious solutions when these matrices become computationally ill-conditioned. There are a couple of notable instances when this can happen. Firstly, when the product of layer thickness with frequency is large, and evanescent waves are to be found, some matrix formulations (e.g. Transfer Matrix) become unstable due to the existence of coefficients that grow and coefficients that decay within the matrices; this is known as the ``large fd" problem \citep{Dunkin1965ComputationFrequencies}. Secondly, matrices can become computationally ill-conditioned when solutions near any of the bulk wavespeeds of a multilayered structure are being searched for. In that scenario, careful filtering needs to be applied to discard such solutions; a task that becomes increasingly difficult when genuine modes exist nearby. Another important caveat of PWRF is that the root finding algorithm sometimes fails to capture a root of the characteristic equation and moves on to searching for the next one. A fine tuning of the convergence parameters of the algorithm fixes this problem, but requires empirical knowledge of the existence of the mode that was missed. This is especially problematic when searching for the complex wavenumber solutions of leaky or attenuating modes. Complex roots are a step change in complexity beyond real roots and are an area in which root-finding routines are known to struggle.

An alternative to PWRF routines are the more recent methods based on a Semi-Analytical Finite Elements (SAFE) approach. An early popular work using SAFE was that of Gavric \citep{Gavric1995ComputationTechnique} on the study of waves propagating in a free rail. In that study, Hamilton’s principle was applied for the derivation of the differential equations of motion while Finite Elements (FE) were used for the discretisation of the waveguide domain and the conversion of the equations of motion into a solvable eigenvalue problem. This allowed for the algebraic retrieval of all the roots of the characteristic equation in the form of eigensolutions. Moreover, SAFE allows for more flexibility when compared to the more traditional PWRF methods, as it can be used for spatially-irregular waveguides and complex structures where the latter becomes unstable. A few examples of this include the use of SAFE for the study of wave propagation in systems of viscoelastic media of arbitrary cross section \citep{Bartoli2006ModelingCross-section}, in axisymmetric damped waveguides \citep{Marzani2008AWaveguides} and in systems of arbitrary cross section coupled to infinite solid media \citep{Castaings2008FiniteMedia}.

In line with the idea of discretising the domain of the structure in question and then obtaining a solution of the dispersion relation in the form of eigenvalues of an algebraic eigenvalue problem, Spectral Collocation Methods (SCM) have also been employed in recent years for the solution of guided wave problems \citep{Adamou2004SpectralMedia}. The use of SCM in physics \citep{Weideman2000ASuite,Trefethen2008SpectralMatlab,Boyd2000ChebyshevMethods}, can be very intuitive and numerically robust, making them an accessible tool for the algebraic solution of differential equations in the form of eigenvalue problems. Also, just like SAFE, and very importantly to our study of leaky waves, SCM guarantee the retrieval of all solutions of the differential equation, including complex valued ones. Most notably, the use of global interpolants, such as polynomials or trigonometric functions, as opposed to local ones, like the piece-wise polynomials used in FE models, makes for an improved convergence rate of discretisation and accuracy of the solution when compared to SAFE methods. 

Adamou and Craster \citep{Adamou2004SpectralMedia} presented solutions for guided waves in simple cylindrical geometries, using SCM with Chebyshev polynomials, demonstrating their potential applications in the field of guided waves. Following that, Karpfinger \textit{et al.} \citep{Karpfinger2008ModelingMethod} successfully used spectral collocation to study multilayered cylindrical structures consisted of isotropic media, Zharnikov \textit{et al.} \citep{Zharnikov2013ApplicationWaveguides} used SCM to study and plot the dispersion curves of an inhomogeneous anisotropic waveguide while Quintanilla \textit{et al.} \citep{Quintanilla2015GuidedMedia,Quintanilla2015ModelingMethod,Quintanilla2016FullMedia} utilised the flexibility of SCM for the study of flat and cylindrical multilayered viscoelastic and generally anisotropic systems.

The studies mentioned above focus on closed waveguide domains, however, the existence of open fluid domain(s) adjacent to the waveguide constitutes a much harder problem, with leaky waves posing a particularly difficult challenge due to the exponential growth of their amplitude far into the open domain(s). While attempting to solve for a fluid-loaded waveguide, root finding techniques on the characteristic equation of Lamb waves \citep{Pavlakovic1997Disperse:Curves,Grabowska1979PropagationSystem,Rokhlin1989OnLayer}, suffer from the aforementioned limitation of missing solutions of PWRF methods, while the often purely complex solutions to be found increase this occurrence. As a result, PWRF can be quite cumbersome when it comes to solving for leaky wave characteristics. 

To overcome the issues of PWRF and complex roots, researchers have recently turned their attention to the discretisation of the domains of their problem in order to obtain algebraic solutions through eigenvalue problems. Mazotti \textit{et al.} \citep{Mazzotti2014UltrasonicValidation} presented a solution in which a SAFE discretisation of the closed waveguide domain was combined with a two and-a-half dimensional boundary element method to represent an unbounded fluid surrounding. To obtain a solution, contour integrals on areas of the complex plane where solutions may lie were performed. It was concluded that, in addition to the a posteriori knowledge of areas of the complex plane where possible solutions might exist, the formulation would be numerically unstable when the guided modes’ poles lie close to the compressional wavespeed in the fluid. Another solution method was presented by Kiefer \textit{et al.} \citep{Kiefer2019CalculatingInteraction} where SCM were implemented for the discretisation of the waveguide domain of a plate loaded on one or both sides by an inviscid fluid, the fluid being identical on both sides in the double-sided case. In that study, interface and boundary conditions were used to account for fluid loading, and a nonlinear eigenvalue problem, in terms of the wavenumber along the length of the waveguide, was produced. Through a change of variables \citep{Hood2017LocalizingApplications}, the resulting nonlinear eigenvalue problem was reduced to a polynomial one, and was later solved by a companion matrix linearisation \citep{Mackey2006StructuredLinearizations} using traditional eigenvalue problem solvers. This method can handle both leaky and trapped modes in a single-sided fluid loaded plate, however, to the best of our knowledge, there is no suitable change of variables to reduce and solve the nonlinear eigenvalue problem that is presented by a waveguide in contact with two different inviscid fluids. 

Alternatively, one could avoid having to deal with a nonlinear eigenvalue problem by effectively discretising the open fluid domain(s) and producing a linear problem. Unfortunately, the difficulties associated with accurately capturing the exponential growth of leaky waves in such a discretision, often forces the derivation of elaborate discritisation techniques such as the use of absorbing layers or absorbing boundary conditions. Fan \textit{et al.} \citep{Fan2008TorsionalFluid}, implemented a SAFE method to model the propagation of leaky waves along an immersed waveguide with arbitrary non-circular cross section. A linear eigenvalue problem was achieved via the discretisation of the fluid domain through the application of an Absorbing Region. Quintanilla \citep{Quintanilla2016PseudospectralMedia} utilised SCM together with a Perfectly Matched Layer to deal with the infinite surrounding medium. The accuracy of such models is, however, considerably dependent on the choice of absorbing layers which in turn are dependent on the nature of the problem and its specific solutions. This means that these choices cannot be made for the general case in advance of performing the solution; a further complicating feature.

Here we overcome the various difficulties outlined earlier and we develop a SCM for the solution of leaky Lamb waves propagating along a waveguide that is in contact with two different inviscid fluid half-spaces. This new solution is applied and demonstrated using a specific example of a flat elastic plate with a different inviscid fluid half-space on each side. However, the solution has powerful generic potential which we believe will enable all classes of leaky waves to be solvable in follow-on developments.

For the discretisation of the closed waveguide domain, Chebyshev polynomials are used. It is worth noting that SAFE could have also been chosen here for the discretisation of the waveguide domain. We opt not to make that choice as the simple geometry of the elastic waveguide does not require the spatial versatility of SAFE, while, on the other hand, it allows for the favourable spectral accuracy of spectral collocation to easily be exploited. To deal with the open fluid half-spaces and the inherent exponential growth of leaky waves, a new technique for the discretisation of the open domains, that is based on Rational Chebyshev polynomials and does not require the use of any absorbing boundaries, is presented here. A polynomial eigenvalue problem is then produced and solved using a companion matrix linearisation and conventional numerical solvers. Finally, the dispersion characteristics and displacement fields of leaky Lamb waves are computed. The case we present here is verified via DISPERSE, to demonstrate the ability of our method to produce accurate results. This is however, without a loss of generality of our method.

This paper is organised as follows - Section \ref{Theory} presents a brief summary of the theory relevant to the solution of the leaky Lamb wave problem of a waveguide in contact with two different inviscid fluid half-spaces. This is followed by a description of the discretisation method in Section \ref{Discretisation}, and a discussion of our results in Section \ref{Results & Discussion}.
\section{Theory}\label{Theory}

\begin{figure}[H]
    \centering
    \includegraphics[scale=0.15]{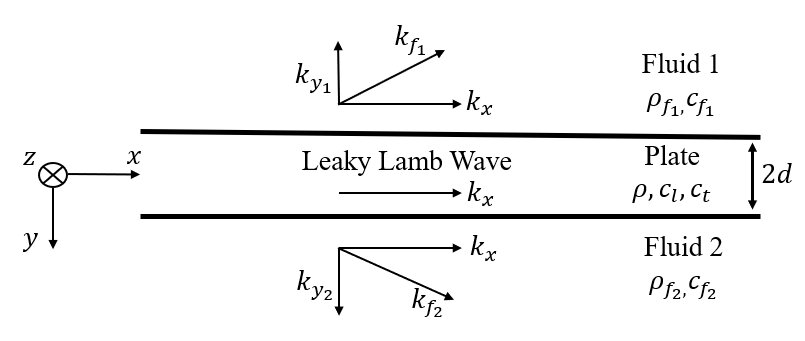}
    \caption{Schematic of an elastic waveguide of thickness $2d$, infinite extent in the $x$ and $z$ directions, density $\rho$, longitudinal wavespeed $c_l$ and transverse wavespeed $c_t$, lying between two inviscid fluids of densities $\rho_{f_1}$, $\rho_{f_2}$ and wavespeeds $c_{f_1}$ and $c_{f_2}$ respectively.}
    \label{Schematic}
\end{figure}

We consider a homogeneous, isotropic, linear-elastic waveguide of infinite extent in the $x$ and $z$ directions, and finite thickness, $2d$, which is immersed between two inviscid fluids. The fluids can be different above, and below, the waveguide and have densities $\rho_{f_1}$, $\rho_{f_2}$ and wavespeeds $c_{f_1}$ and $c_{f_2}$ respectively, while the waveguide has density $\rho$, longitudinal wavespeed $c_l$ and transverse wavespeed $c_t$. A schematic of the cross section of the waveguide, also showing the coordinate axes, is shown in Fig. \ref{Schematic}. 

The centreline of the waveguide is along $y=0$, and the axis orientation has waves propagating in the positive $x$ direction with wavenumber $k_x$. For given angular frequency, $\omega$, the phase velocity is given by $c_{ph}={\omega}/\Re(k_x)$ and attenuation in the propagation direction is $\alpha_x=\Im(k_x)$, with $\Re(\boldsymbol{\cdot})$ and $\Im(\boldsymbol{\cdot})$ denoting real and imaginary components of a complex number . Waves with displacements in the $z$ direction decouple from those with displacements in the $x$ and $y$ directions \citep{Quintanilla2017TheWaveguides}; the latter encompass the leaky Lamb waves that we aim to determine and hence we only require displacements $\bar{u}_x, \bar{u}_{x_{f_1}}$ and $\bar{u}_{x_{f_2}}$ and $\bar{u}_y, \bar{u}_{y_{f_1}}$ and $\bar{u}_{y_{f_2}}$. 

We denote with $\boldsymbol{\bar{u}}$ the vector displacement in the waveguide 
\begin{equation}\label{displacement vector solid} 
        \boldsymbol {\bar{u}}=\begin{pmatrix}
\bar{u}_x \\
\bar{u}_y \\
0 \end{pmatrix},
\end{equation}
and by $\boldsymbol{\bar{u}}_{f_1}$,  $\boldsymbol{\bar{u}}_{f_2}$ the vector displacements in the two fluids. By employing a Helmholtz decomposition \citep{Achenbach1973WaveSolids,Rose2014UltrasonicMedia}, they can be written as
\begin{equation} \label{displacement vectors fluids} 
  \boldsymbol {\bar{u}}_{f_1}=\begin{pmatrix}
\bar{u}_{x_{f_1}} \\
\bar{u}_{y_{f_1}} \\
0 \end{pmatrix}=\nabla \bar{\phi}_{f_1}, 
  \hspace{0.5cm} \boldsymbol {\bar{u}}_{f_2}=\begin{pmatrix}
\bar{u}_{x_{f_2}} \\
\bar{u}_{y_{f_2}} \\
0 \end{pmatrix}=\nabla \bar{\phi}_{f_2},
\end{equation}

\noindent with $\bar{\phi}_{f_1}$ and $\bar{\phi}_{f_2}$ the displacement potentials in the two fluids. The quantities $\bar{u}_x$, $\bar{u}_y$, $\bar{\phi}_{f_1}$ and $\bar{\phi}_{f_2}$ are functions of $x,y$ and $t$. We are assuming plane harmonic waves in the form 
\begin{equation}
\label{u_x} \bar{u}_x(x,y,t)=u_x(y)\exp(i(k_xx-\omega t)),
\end{equation}

\noindent with the common exponential factor henceforth omitted and considered understood, and similarly for $\bar{u}_y(x,y,t), \bar{\phi}_{f_1}(x,y,t), \bar{\phi}_{f_2}(x,y,t)$. 

The two dimensional wave motion in the isotropic solid is governed by the momentum equations \citep{supplementary}

\begin{align}
\label{equation of motion solid x}
    \rho \omega^2 u_x-(\lambda+2\mu)k_x^2u_x+(\lambda+\mu) ik_x\frac{\partial u_y}{\partial y}+\mu\frac{\partial^2 u_x}{\partial y^2}&=0, \\
\label{equation of motion solid y}
    \rho \omega^2 u_y-\mu k_x^2u_y+(\lambda+\mu) ik_x\frac{\partial u_x}{\partial y}+(\lambda+2\mu)\frac{\partial^2 u_y}{\partial y^2}&=0,
\end{align}

\noindent with $\lambda$, $\mu$ the elastic Lam\'e parameters. We follow Nayfeh and Nagy \citep{Nayfeh1997ExcessLoading} for the fluids, specialising their viscous model to be inviscid, to obtain  fluid momentum equations \citep{supplementary} 
\begin{equation}
\label{equation of motion fluids}
    \rho_{f_j} \omega^2 \phi_{f_j}-\lambda_{f_j}k_x^2\phi_{f_j}+\lambda_{f_j}\frac{\partial^2 \phi_{f_j}}{\partial y^2}=0,
\end{equation}

\noindent with the subscripts $j=1,2$ corresponding to the fluids above and below the waveguide respectively, and $\lambda_{f_j}$ the Lam\'e parameter of each fluid.

To fully describe wave motion in the immersed waveguide system, in addition to Eqs. \eqref{equation of motion solid x}-\eqref{equation of motion fluids}, interface conditions are required. For a two-sided fluid-loaded plate, these are continuity of traction and of normal displacements at the two fluid-solid interfaces \citep{Rose2014UltrasonicMedia}; for interfaces located at $y_1=-d$ and $y_2=d$, the traction conditions translate to $(\sigma_{y_{f_j}}-\sigma_y)|_{y=y_j}=0$ and $\sigma_{xy}|_{y=y_j}=0$, and the displacement conditions to $(u_{y_{f_j}}-u_y)|_{y=y_j}=0$.  
These are explicitly expressed \citep{supplementary} as 

\begin{align}
\label{continuity of normal stress} 
   \left. \left[(\lambda+2\mu)\frac{\partial u_y}{\partial y}+\lambda ik_xu_x+\rho_{f_j}\omega^2\phi_{f_j}\right]\right|_{y=y_j}=0,\\
\label{zero shear} 
   \left. \left[ik_xu_y+\frac{\partial u_x}{\partial y}\right]\right|_{y=y_j}= 0,\\
    \label{continuity of normal displacements}
    \left.\left[\frac{\partial \phi_{f_j}}{\partial y}-u_y \right]\right|_{y=y_j}=0.
\end{align}

Wave propagation in the waveguide is captured by a single wavenumber, $k_x$, whereas the inhomogeneous bulk waves that propagate in the fluids, at an angle from the waveguide, see Fig. \ref{Schematic}, require additional wavenumbers; these represent partial waves propagating in the transverse, $y$ direction. The aforementioned wavenumbers are denoted by $k_{y_1}$ and $k_{y_2}$ for the two fluids. Similar to $k_x$, the real parts of these transverse wavenumbers describe the speed of propagation of the waves in the $y$ direction and their imaginary parts describe their attenuation or growth as they propagate along the same direction.  
We define the propagation wave vectors of these inhomogeneous bulk waves as 

\begin{equation} \label{wavevectors}
    \boldsymbol{k_{f_1}}:=\begin{pmatrix}
    k_x \\
    k_{y_1}
    \end{pmatrix} \text{ and }
    \boldsymbol{k_{f_2}}:=\begin{pmatrix}
    k_x \\
    k_{y_2}
    \end{pmatrix}.
\end{equation}

For inviscid fluids, the magnitude of these wave vectors is real and is given by 
\begin{equation}
    \label{wavevectors_magnitude}
     k_{f_j}:=|\boldsymbol{k_{f_j}}|=\frac{\omega}{c_{f_j}},\text{ for }j=1,2,
\end{equation}

\noindent and the following trigonometric relations hold 
\begin{equation}
\label{trigonometric_identity wavevector}
k_{f_j}^2=k_x^2+k_{y_j}^2,\text{ for }j=1,2.   
\end{equation}

Consequently, the longitudinal wave potentials, $\phi_{f_1}$ and $\phi_{f_2}$, in the two fluid half-spaces, are \citep{Declercq2005TheWaves,Nayfeh1997ExcessLoading} 

\begin{align}
    \label{fluid wave potentials 1}
    &\phi_{f_1}(y)=A_1\exp({-ik_{y_1}(y+d)}), \\
    \label{fluid wave potentials 2}
    &\phi_{f_2}(y)=A_2\exp({ik_{y_2}(y-d)}),
\end{align}
where $A_1, A_2$ are constants. 

One approach to formulating the problem mathematically is  to substitute Eqs. \eqref{fluid wave potentials 1}, \eqref{fluid wave potentials 2} in Eqs. \eqref{equation of motion fluids}-\eqref{continuity of normal displacements}, discretise the waveguide and arrive at a nonlinear eigenvalue problem \citep{Kiefer2019CalculatingInteraction}. For a single fluid, either on one side or on both sides, and leaky Lamb waves in a waveguide, Kiefer \textit{et al.} \citep{Kiefer2019CalculatingInteraction}  employ a spectral collocation method using Chebyshev polynomials to discretise the waveguide; a neat change of variables \citep{Hood2017LocalizingApplications} reduces the algebraic nonlinear eigenvalue problem to a polynomial eigenvalue problem which is then solvable with well-established methods. We are aiming at a general approach for leaky waves in immersed waveguides or pipes, while fluid coupling of even straight waveguides, by two different fluids, leads to a more complicated nonlinear eigenvalue problem that appears not to reduce quite so pleasantly. We employ a spectral collocation discretisation of both the waveguide and of the two open spaces of the inviscid fluids; the key being the discretisation of the fluids which we describe in detail in the following section. This discretisation results in a polynomial eigenvalue problem with the complex wavenumber $k_x$ as its eigenvalue and the wave displacements, complemented by the wave amplitudes of Eqs. \eqref{fluid wave potentials 1} and \eqref{fluid wave potentials 2}, as the eigenvector; this deals with the exponential growth of the leaky modes without the need to use any absorbing or perfectly matched layers in the fluid domains and leads to an eigenvalue problem that can be solved with standard methods.

\section{Discretisation} \label{Discretisation}
The discretisation of both the waveguide and the open fluid domains is done using spectral collocation \citep{Fornberg1996AMethods,Trefethen2008SpectralMatlab}.

In brief, spectral collocation revolves around creating differentiation matrices: consider a continuous function $g(y)$, with $y$ in some domain $\mathcal{C}$, whose derivative is to be approximated. A set of distinct interpolation points $\{y_i\}_{i=1}^N\in \mathcal{C}$, for some $N\in\mathbb{N}$, also known as collocation points \citep{Fornberg1996AMethods,Boyd2000ChebyshevMethods}, 
are fixed and then $\boldsymbol{g}$, the vector of $g$ evaluated at the collocation points,  and $\boldsymbol{g}^{(l)}$, the vector of their $l^{th}$ derivatives, are related via the $l^{th}$ differentiation matrix of $g$, $D^{(l)}$, such that
\begin{equation}
    \boldsymbol{g}^{(l)} \approx D^{(l)}\boldsymbol{g}.
\end{equation}

 The identification of the collocation points is the key step, and a common choice, when $\mathcal{C}$ is a bounded interval and $g$ is a smooth function, is that of Chebyshev collocation points for which the Chebyshev differentiation matrices exhibit spectral accuracy \citep{Fornberg1996AMethods,Boyd2000ChebyshevMethods}. This accuracy, together with their flexibility and established use in guided wave problems \citep{Kiefer2019CalculatingInteraction,Adamou2004SpectralMedia,Quintanilla2015GuidedMedia}, motivates our use of them for the waveguide region. To treat the open spaces of fluid, we use a generalisation of Chebyshev methods - specifically of rational Chebyshev \citep{Boyd2000ChebyshevMethods} approaches; these have been used for guided waves in optics, i.e. in the modal analysis of a multilayered optical waveguide \citep{Abdrabou2016EfficientAnalysis}. Rational Chebyshev nodes arise from the mapping of Chebyshev collocation nodes onto an open space, infinite or semi-infinite, and preserve the performance benefits of traditional Chebyshev differentiation matrices.  We use the differentiation matrix suite provided by Weideman and Reddy \citep{Weideman2000ASuite} to obtain the first and second order differentiation matrices,  denoted by $D^{(1)}$ and $D^{(2)}$ respectively, using $N$ Gauss–Lobatto Chebyshev collocation points \begin{equation}
    \label{chebyshev collocation points}
    y_i=\cos\left(\frac{(i-1)\pi}{N-1}\right), 
\end{equation}

\noindent for $i=1,\dots,N$ and $y_i\in[-1,1]$.

The treatment of the waveguide follows \citep{Adamou2004SpectralMedia}: multiplication by $d$ maps the $N$ collocation points of Eq. \eqref{chebyshev collocation points} onto the bounded interval of the waveguide, $[-d,d]$; the collocation points in that interval are $\{y_{p}^{i}\}_{i=1}^{N}$. The corresponding differentiation matrices, approximating the first and second order derivatives in the solid domain, are $D_{p}^{(1)}=\left(\frac{1}{d}\right)D^{(1)}$ and $D_{p}^{(2)}=\left(\frac{1}{d^2}\right)D^{(2)}$.

To treat the fluid domains, and build in the exponential growth of leaky waves, we follow a different approach. Consider the complex variable mappings $h_{f_1}:[-1,1]\longrightarrow \mathbb{C}$ and $h_{f_2}:[-1,1]\longrightarrow \mathbb{C}$ given by 

\begin{align}
\label{h1 map}
    h_{f_1} : y \longrightarrow -d-\zeta_{f_1}\frac{1-y}{1+y}, \\
\label{h2 map}
    h_{f_2} : y \longrightarrow d+\zeta_{f_2}\frac{1+y}{1-y},
\end{align} 

\noindent with $\zeta_{f_1},\zeta_{f_2}\in \mathbb{C}$; the choice of the two complex parameters $\zeta_{f_1}$ and $\zeta_{f_2}$ plays an important role in our discretisation of the fluid domains as importantly we are mapping a real interval to a semi-infinite interval in complex space. To guide their choice we note that the wave potentials in Eqs. \eqref{fluid wave potentials 1} and \eqref{fluid wave potentials 2}, for exponential growth of the outgoing leaky waves, require the transverse wavenumbers $k_{y_1}$ and $k_{y_2}$ to lie in the 4$^{th}$ complex quadrant.

The key step is the mapping to complex space, and not to a real domain: A choice in previous work of $\zeta_{f_1},\zeta_{f_2}\in \mathbb{R}_{>0}$, i.e. real, as used in, say, the optics waveguide \citep{Abdrabou2016EfficientAnalysis}, maps the collocation points of Eq. \eqref{chebyshev collocation points}, through  $h_{f_1}$ and $h_{f_2}$, to $[-\infty,-d]$ and $[d,\infty]$ respectively. The implicit exponential growth of leaky waves prevents the QZ-algorithm of conventional eigenvalue solvers from reliably retrieving the solution and requires the use of perfectly matched, absorbing layers or outgoing wave boundary conditions at infinity which simply transfers the numerical problems normally encountered to elsewhere in the problem. 

Instead, we take $\zeta_{f_1},\zeta_{f_2}\in \mathbb{C}$, i.e. complex, and to lie in the 1$^{st}$ complex quadrant. This choice has  $h_{f_1}$ mapping the collocation points of Eq. \eqref{chebyshev collocation points} to a path in the 3$^{rd}$ complex quadrant, with the point at $1$ being mapped to the interface at $y=-d$. Similarly, $h_{f_2}$ maps the collocation points to a path in the 1$^{st}$ complex quadrant, with the point at $-1$ being mapped to the interface at $y=d$. These result in a discretisation of two paths in the complex quadrants, which we denote by $\mathbb{C}_{f_1}$ and $\mathbb{C}_{f_2}$ respectively. 

The analytic continuations of each of the fluid wave potentials of Eqs. \eqref{fluid wave potentials 1} and \eqref{fluid wave potentials 2}, with domains of definition the complex paths $\mathbb{C}_{f_j}$ are henceforth referred to as the complex fluid wave potentials and are denoted by $\tilde{\phi}_{f_j}$. Both of $\tilde{\phi}_{f_1}$ and $\tilde{\phi}_{f_2}$ still satisfy the momentum equations of Eq. \eqref{equation of motion fluids} for the two fluids and all six of the interface conditions of Eqs. \eqref{continuity of normal stress}-\eqref{continuity of normal displacements}. Importantly,  choices of $\zeta_{f_1},\zeta_{f_2}$, make the numerical problem have exponential decay, even for leaky modes, while simultaneously attaining the physical solution on the two fluid-solid interfaces, as the choice of mapping guarantees a collocation point on the real space of those interfaces; the inhomogeneous plane waves in the two fluids mean that values at the interfaces fully determine motion in each fluid, hence solving the complex valued problem arising from using $h_{f_1}$ and $h_{f_2}$, gives the physical solution for the leaky waves.

Using the mappings we can create differentiation matrices, taking $N_{f_1}$ and $N_{f_2}$ as the number of collocation points for each of the fluids, we set $\{y_{f_1}^{m}\}_{m=1}^{N_{f_1}}$ the collocation points in $\mathbb{C}_{f_1}$ and  $\{y_{f_2}^{n}\}_{n=1}^{N_{f_2}}$ the collocation points in $\mathbb{C}_{f_2}$; repeated use of the chain rule gives the differentiation matrices for the fluids in the two complex domains. The $N_{f_1}\times N_{f_1}$ differentiation matrices for the fluid at the top of the plate are given by 

\begin{align}
    \label{first order differentiation matrix fluid 1}
    D_{f_1}^{(1)}     =&\text{diag}\left( \frac{2\boldsymbol{\zeta}_{f_1}}{\lbrack \boldsymbol{\zeta}_{f_1}-(\boldsymbol{y}_{f_1}+\boldsymbol{d})\rbrack^2}\right)D^{(1)}, \\
     \label{second order differentiation matrix fluid 1}
     \begin{split} 
     D_{f_1}^{(2)}      =&\text{diag}\left(\frac{4\boldsymbol{\zeta}_{f_1}}{\lbrack \boldsymbol{\zeta}_{f_1}-(\boldsymbol{y}_{f_1}+\boldsymbol{d})\rbrack^3}\right)D^{(1)} \\ &+\text{diag}\left(\frac{4\boldsymbol{\zeta}_{f_1}^2}{\lbrack \boldsymbol{\zeta}_{f_1}-(\boldsymbol{y}_{f_1}+\boldsymbol{d})\rbrack^4}\right)D^{(2)},
     \end{split}
\end{align}

\noindent and the $N_{f_2}\times N_{f_2}$ ones for the fluid at the bottom of the plate are given by 

\begin{align}
 \label{first order differentiation matrix fluid 2}
    D_{f_2}^{(1)}=&\text{diag}\left( \frac{2\boldsymbol{\zeta}_{f_2}}{\lbrack \boldsymbol{\zeta}_{f_2}+(\boldsymbol{y}_{f_2}-\boldsymbol{d})\rbrack^2}\right)D^{(1)}, \\
         \label{second order differentiation matrix fluid 2}
     \begin{split}
     D_{f_2}^{(2)}=&\text{diag}\left(\frac{-4\boldsymbol{\zeta}_{f_2}}{\lbrack \boldsymbol{\zeta}_{f_2}+(\boldsymbol{y}_{f_2}-\boldsymbol{d})\rbrack^3}\right)D^{(1)}\\
     &+\text{diag}\left(\frac{4\boldsymbol{\zeta}_{f_2}^2}{\lbrack \boldsymbol{\zeta}_{f_2}+(\boldsymbol{y}_{f_2}-\boldsymbol{d})\rbrack^4}\right)D^{(2)}.
     \end{split}
\end{align}

\noindent Here, $\boldsymbol{y}_{f_1}$ and $\boldsymbol{y}_{f_2}$ are vectors containing the collocation points for the two fluids, $\boldsymbol{d}=(d,\dots,d)^T$ is a column vector of appropriate size, and likewise for $\boldsymbol{\zeta}_{f_1}$ and $\boldsymbol{\zeta}_{f_2}$. The diagonal matrices in Eqs. \eqref{first order differentiation matrix fluid 1}-\eqref{second order differentiation matrix fluid 2} are the matrices whose diagonals are the expressions in parenthesis, evaluated at each of the collocation points, and have zeros everywhere else. 

Due to our choice of discretisation, now numerically  extracting a decaying function rather than an exponentially growing one, the choice of boundary condition at infinity is relevant only insofar as we want to ensure the solution tends to zero at infinity; this is achieved by taking Dirichlet zero boundary conditions at the farthest collocation point in the complex space. This is applied by deleting the corresponding rows and columns of our differentiation matrices, as well as the corresponding collocation point. For the top fluid, we delete $y_{f_1}^1$ and the first rows and columns of $D_{f_1}^{(1)}$ and  $D_{f_1}^{(2)}$ and for the bottom fluid, we delete $y_{f_2}^{N_{f_2}}$ and the last rows and columns of $D_{f_2}^{(1)}$ and  $D_{f_2}^{(2)}$. The $x$ and $y$ displacements in the waveguide, each evaluated at the $N$ collocation points within the domain of the solid, are given in vector form by $\boldsymbol{u_x}=\begin{pmatrix}u_x(y_p^{1}) \\
\vdots \\
u_x(y_p^{N})
\end{pmatrix}$ and $\boldsymbol{u_y}=\begin{pmatrix}
u_y(y_p^{1}) \\
\vdots \\
u_y(y_p^{N})
\end{pmatrix}$ while the complex fluid wave potentials evaluated at the collocation points of each of the fluids are given by $\boldsymbol{\tilde{\phi}}_{f_1}=\begin{pmatrix} \tilde{\phi}_{f_1}(y_{f_1}^{2}) \\
\vdots \\
\tilde{\phi}_{f_1}(y_{f_1}^{N_{f_1}})\end{pmatrix}$ and $\boldsymbol{\tilde{\phi}}_{f_2}=\begin{pmatrix} \tilde{\phi}_{f_2}(y_{f_2}^{1}) \\
\vdots \\
\tilde{\phi}_{f_2}(y_{f_2}^{N_{f_2}-1})\end{pmatrix}$.

We can now proceed to generate the eigenvalue problem we wish to solve. We start by representing the differential operators of Eqs. \eqref{equation of motion solid x}-\eqref{equation of motion fluids}, using the differentiation matrices $D_p^{(l)}$ and $D_{f_j}^{(l)}$, for $j=1,2$ and $l=1,2$,  as the matrix operators 

\begin{align}
\label{differential operator plate}
L_p&=\begin{pmatrix} L_{p1} \\
L_{p_2}
\end{pmatrix}, \\ 
 \label{differential operator fluids}
    L_{f_j}&=k_x^2(-\lambda_{f_j}I)+\rho_{f_j} \omega^2 I+\lambda_{f_j}D_{f_j}^{(2)},
\end{align} 

\noindent with 

\begin{equation}
    \label{differential operator plate 1}
    \begin{split}
        L_{p_1} =&k_x^2\begin{bmatrix}
-(\lambda+2\mu)I &0 \end{bmatrix}+
k_x\begin{bmatrix}
0&i(\lambda+\mu)D_p^{(1)}
\end{bmatrix}\\
&+\begin{bmatrix}\rho \omega^2I+\mu D_p^{(2)}&0 \\
\end{bmatrix},
    \end{split}
\end{equation}

\noindent and 

\begin{equation}
\label{differential operator plate 2}
\begin{split}
    L_{p_2}=&k_x^2\begin{bmatrix} 0&-\mu I \end{bmatrix}+k_x\begin{bmatrix} i(\lambda+\mu)D_p^{(1)} &0 \end{bmatrix} \\
     &+ \begin{bmatrix}0& \rho \omega^2I +(\lambda+2\mu)D_p^{(2)}\end{bmatrix}.
\end{split}
\end{equation}

In Eqs. \eqref{differential operator fluids}-\eqref{differential operator plate 2}, $I$ is the identity matrix of appropriate size;  these matrix operators allow us to write the equations of motion of the immersed plate problem as the eigenvalue problem 
\begin{equation}
    \label{eigenvalue problem before boundary}
\Tilde{L}\boldsymbol{u}=\boldsymbol{0},
\end{equation}

\noindent with $\Tilde{L}=\begin{pmatrix}L_{f_1} & 0&0\\
0& L_{p}  & 0 \\
0 & 0& L_{f_2} 
\end{pmatrix}$
and $\boldsymbol{u}=\begin{pmatrix} \boldsymbol{\tilde{\phi}}_{f_1} \\
\boldsymbol{u_x} \\
\boldsymbol{u_y} \\
\boldsymbol{\tilde{\phi}}_{f_2}\end{pmatrix}$.
In Eqs. \eqref{differential operator plate}-\eqref{differential operator plate 2}, we separate the matrix operators into  $k_x^2$-dependent, $k_x$-dependent and constant components; this is convenient for the polynomial eigenvalue problem formulation we use later. Similarly, we separate $\Tilde{L}$ with $\Tilde{L}_2, \Tilde{L}_1$ and $\Tilde{L}_0$ being the matrices for the $k_x^2$-dependent, $k_x$-dependent and constant components respectively: Eq. \eqref{eigenvalue problem before boundary} is rewritten as
\begin{equation}
    \label{eigenvalue problem before boundary polynomialy split}
    \left(k_x^2\Tilde{L}_2+k_x\Tilde{L}_1+\Tilde{L}_0\right)\boldsymbol{u}=0.
\end{equation}

The final step is to incorporate the interface conditions Eqs. \eqref{continuity of normal stress}-\eqref{continuity of normal displacements}. To achieve this, we firstly write each of those interface conditions as rows of two $3\times (N_{f_1}+2N+N_{f_2}-2)$ matrices. For simplicity, we represent those with two $3\times 4$ matrices. Their first and last columns correspond to $\boldsymbol{\tilde{\phi}}_{f_1}$ and $\boldsymbol{\tilde{\phi}}_{f_2}$, whereas the second and third columns correspond to $\boldsymbol{u_x}$ and $\boldsymbol{u_y}$. The matrix of interface conditions at the top interface is

\begin{equation} \label{interface at y=-d}
\begin{split}
    IC_{-d}=&\left. k_x\begin{bmatrix}
    0 & i\lambda I & 0 & 0 \\
    0 & 0 & i I & 0 \\
    0 & 0 & 0 & 0
    \end{bmatrix}\right\vert_{y=-d} \\
     &+ \left.\begin{bmatrix}
    \rho_{f_1}\omega^2 I & 0 & (\lambda+2\mu)D_p^{(1)} & 0 \\
    0 &  D_p^{(1)} & 0 & 0 \\
     D_{f_1}^{(1)} & 0 & -I & 0
    \end{bmatrix}\right\vert_{y=-d},
\end{split}
\end{equation}

\noindent where the elements are row vectors, representing the discrete evaluation of the differential operators of the interface conditions at the interface $y=-d$. This evaluation at $y=-d$, corresponds to picking the rows of the differential operators appearing in Eq. \eqref{interface at y=-d} that correspond to the collocation point at that interface. By our choice of discretisation, for the complex fluid potential of the fluid at the top of the plate, we need to pick the $(N_{f_1}-1)^{\text{th}}$ row of the matrices whereas for the plate displacements, we need to pick the $1^{st}$ rows of the matrices.

Similarly, the matrix of interface conditions at the bottom interface is

\begin{equation} \label{interface at y=d}
\begin{split}
    IC_{d}=&k_x \left.\begin{bmatrix}
    0 & i\lambda I & 0 & 0 \\
    0 & 0 & i I & 0 \\
    0 & 0 & 0 & 0
    \end{bmatrix}\right\vert_{y=d} \\
    &+\left.\begin{bmatrix}
    0 & 0 & (\lambda+2\mu)D_p^{(1)} & \rho_{f_2}\omega^2I  \\
    0 & D_p^{(1)} & 0 & 0 \\
     0 & 0 & -I & D_{f_2}^{(1)} 
    \end{bmatrix}\right\vert_{y=d}.
\end{split}
\end{equation}

\noindent The same convention for the evaluation of the differential operators is used here. For the evaluation at the interface $y=d$, we need to pick the $1^{st}$ row of the operators for the complex fluid potential of the fluid at the bottom of the plate and the $N^{\text{th}}$ row of the operators for the plate displacements.

Secondly, we incorporate those interface conditions in $\Tilde{L}$ by replacing the rows of $\Tilde{L}$ that correspond to the two interfaces, with rows from the interface conditions, as those were described in Eqs. \eqref{interface at y=-d} and \eqref{interface at y=d}. Again, due to our discretisation and construction of $\Tilde{L}$ from the matrix differential operators of Eqs. \eqref{differential operator plate} and \eqref{differential operator fluids}, the rows of $\Tilde{L}$ corresponding to the interface $y=-d$ are $N_{f_1}-1,  N_{f_1}$ and $N_{f_1}+N$. We replace each of those with a row from Eq. \eqref{interface at y=-d}. Similarly, the rows of $\Tilde{L}$ corresponding to the interface at $y=d$ are $N_{f_1}+N-1,N_{f_1}+2N-1$ and $N_{f_1}+2N$, and we replace each of those with a row from Eq. \eqref{interface at y=d}. Finally, we arrive at a matrix, $L$, with an analogous decomposition to the one of Eq. \eqref{eigenvalue problem before boundary polynomialy split}. The resulting polynomial eigenvalue problem is

\begin{equation}
    \label{polynomial eigenvalue problem}
    \left(k_x^2L_2+k_xL_1+L_0\right)\boldsymbol{u}=\boldsymbol{0},
\end{equation}

\noindent where for a given $\omega$, $k_x$ is our eigenvalue and $\boldsymbol{u}$ is our eigenvector. 
To solve the polynomial eigenvalue problem, we employ companion matrix linearisation \citep{Mackey2006StructuredLinearizations} by setting $\boldsymbol{U}=[k_x\boldsymbol{u},\boldsymbol{u}]^T$ and define the square companion matrices 

\begin{align}
A&=\begin{bmatrix}
-L_1 & -L_0 \\
I & 0
\end{bmatrix}, \\
B&=\begin{bmatrix}
L_2 & 0 \\
0 & I
\end{bmatrix},
\end{align} 

\noindent with $I$ the identity matrix of dimension $(N_{f_1}+2N+N_{f_2}-2) \times (N_{f_1}+2N+N_{f_2}-2)$. Then, the eigenvalue problem of Eq. \eqref{polynomial eigenvalue problem} is equivalent to the generalised linear eigenvalue problem,

\begin{equation}
    \label{final eigenvalue problem}
    (A-k_xB)\boldsymbol{U}=\boldsymbol{0},
\end{equation}

 \noindent which is solvable with conventional eigenvalue solvers, with eigenvalue $k_x$ and eigenvector $\boldsymbol{U}$. 

\section{Results \& Discussion}\label{Results & Discussion}
The leaky wave SCM is now employed to find the dispersion curves of the leaky modes for an elastic waveguide with different inviscid fluids above and below the waveguide;  these curves are validated  with the commercially available software DISPERSE \citep{Pavlakovic1997Disperse:Curves} as are the displacements obtained as the eigenvectors of the leaky SCM.  DISPERSE is a well-established software designed for the generation of dispersion curves in multilayered media, embedded or in vacuo, using matrix techniques and a PWRF algorithm \citep{Pavlakovic1997Disperse:Curves,Lowe2013UsersCurves} and it provides a good basis for comparison with our method as its reliability has been demonstrated in multiple scenarios \citep{Diligent2003PredictionIncident,Vogt2004MeasurementWaves,Cegla2005MaterialSensor,Ma2007FeasibilityWaves,Leinov2016UltrasonicPipes}.

We choose a typical example and compute the dispersion curves of a $1$mm thick brass plate ($\rho=8.4$ g/$\mbox{cm}^3$, $c_t=2.2$ m/ms, $c_l=4.4$ m/ms) loaded on the top side by water ($\rho_{f_1}=1$ g/cm$^3$, $c_{f_1}=1.5$ m/ms) and at the bottom side by diesel oil ($\rho_{f_2}=0.8$ g/cm$^3$, $c_{f_2}=1.25$ m/ms). For the choice of complex parameters $\zeta_{f_j}$, we draw upon  \citep{supplementary}. For a complex wavenumber solution $k_x$, and corresponding transverse wavenumbers $k_{y_j}$, a choice of $\zeta_{f_j}$ such that 

\begin{equation} \label{choice of complex parameters}
    \Im(\zeta_{f_j})>-\Re(\zeta_{f_j})\frac{\Im(k_{y_j})}{\Re(k_{y_j})},
\end{equation}

\noindent results in exponentially decaying potentials on the complex paths $\mathbb{C}_{f_j}$. Without loss of generality, we make the choice of $\Re(\zeta_{f_j})=1$, maintaining in this way the span of the Gauss–Lobatto Chebyshev collocation points of Eq. \eqref{chebyshev collocation points}, along the complex paths $\mathbb{C}_{f_j}$; we can without loss of generality choose $\Im(\zeta_{f_j})>-{\Im(k_{y_j})}/{\Re(k_{y_j})}$. 

The choice can be automated and we use the following rule-of-thumb: we first solve for the same waveguide immersed in two identical inviscid fluids \citep{Kiefer2019CalculatingInteraction}. This is done twice, solving for a brass waveguide fully immersed in water and then in oil, each time recording the maximum value of the ratio $-{\Im(k_{y})}/{\Re(k_{y})}$ for leaky wave solutions in the frequency range of interest of our problem. These maximum ratios provide initial guesses for the ratios $-{\Im(k_{y_j})}/{\Re(k_{y_j})}$;  in the angular frequency range 0-20 MHz, those were found to be less than 1. 

In light of this, for the computation of dispersion curves with our SCM, we take the conservative choice of $\zeta_{f_j}=1+10i$, satisfying in this way Eq. \eqref{choice of complex parameters}. As this translates to a rapidly decaying solution, we also choose a sufficient number of collocation points to capture this decay. A conservative choice is $N=N_{f_1}=N_{f_2}=30$, which yields a convergent solution. We have investigated the stability of the algorithm regarding the choices of $\zeta_{f_j}$ and the numbers of collocation points and the scheme is very robust to the choices taken. All  results are obtained on a computer with 32GB RAM memory and an i7-10700 CPU. 

\begin{figure}[h]
\centering
\includegraphics[scale=0.40]{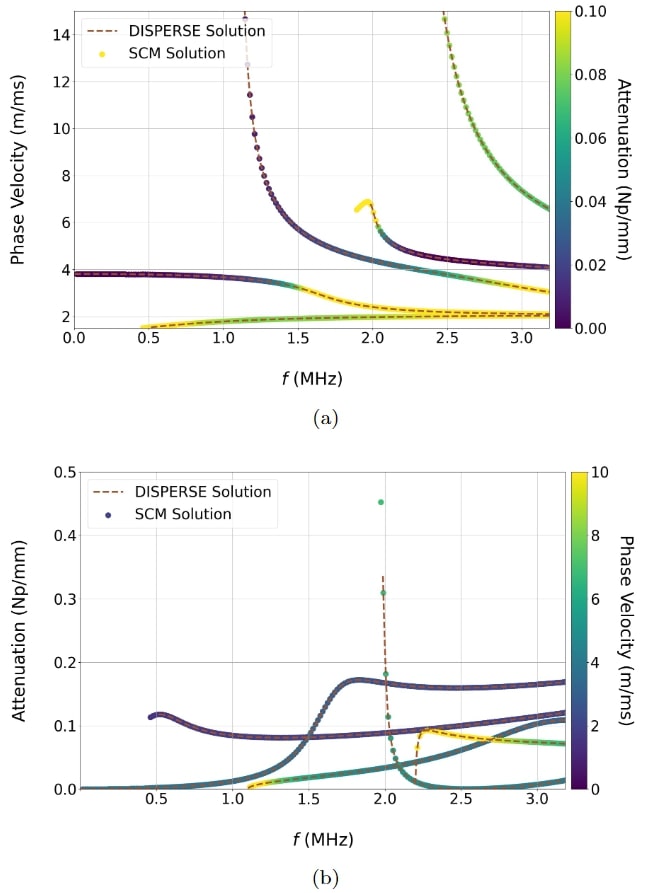}
\caption{Comparison of dispersion curves obtained from our SCM and from DISPERSE, for a 1mm thick brass plate, loaded on one side by water and on the other by diesel oil. Fig. \hyperref[Dispersion Curves]{\ref{Dispersion Curves}a} plots phase velocity against frequency, while Fig. \hyperref[Dispersion Curves]{\ref{Dispersion Curves}b} plots attenuation against frequency.}
\label{Dispersion Curves}
\end{figure}

The SCM results are compared with those from DISPERSE in Fig. \ref{Dispersion Curves} with pleasing agreement for both the phase velocity and attenuation. There are infinitely many possible modes that can propagate through the waveguide and we have performed filtering  prior to plotting so we illustrate those modes of most physical interest. The filtering  comprises a positively propagating wave requirement with $\Re(k_x)>0$ and attenuation with $\Im(k_x)>0$ and $|k_x|\neq \infty$. For plotting purposes we took an arbitrary choice of cut-off attenuation at 1 Np/mm. 

\begin{figure}[t]
\centering
\includegraphics[scale=0.44]{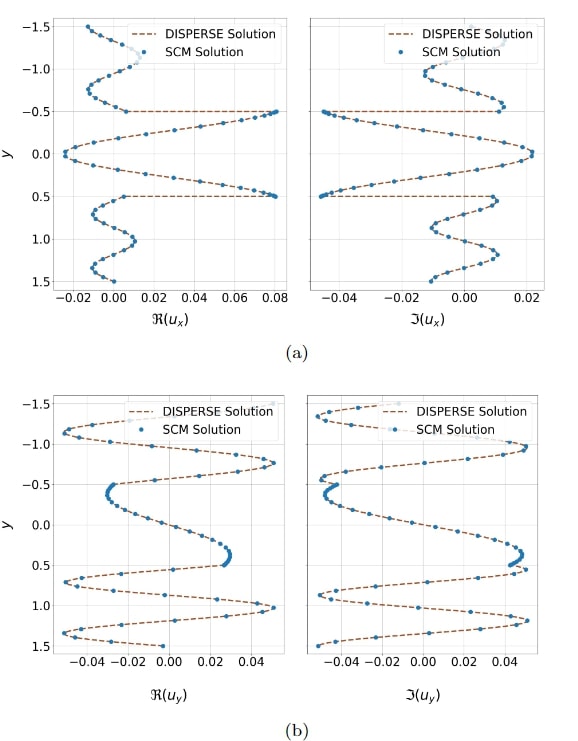}
\caption{Mode shapes displaying the $y$-dependent component of displacements for a 1mm thick brass plate loaded on one side by water and on the other by diesel oil, for the frequency-phase velocity pair $(f,c_{ph})\approx(2.02, 6.18)$. Fig. \hyperref[Mode Shapes]{\ref{Mode Shapes}a} plots the real and imaginary parts of $u_x(y)$ and Fig. \hyperref[Mode Shapes]{\ref{Mode Shapes}b} plots the real and imaginary parts of $u_y(y)$. In both plots, the results from DISPERSE and the scaled results from our SCM are presented.}
\label{Mode Shapes}
\end{figure}

The SCM conveniently retrieves the values of the amplitudes of the fluid potential and the displacements in the solid from the eigenvector through $\boldsymbol{U}$;   as these fully describe the acoustic fields in the fluids and the solid, we  get the whole of the wave field without any extra effort. The displacements in the fluids are  computed using
the Helmholtz decomposition of the fluid displacements in Eq. \eqref{displacement vectors fluids}, and the wave potentials in Eqs. \eqref{fluid wave potentials 1} and \eqref{fluid wave potentials 2}. Using the water-brass-oil system of Fig. \ref{Dispersion Curves}, we compare the $y$-dependent component of the displacements, in both the direction of propagation and the normal direction, with the corresponding DISPERSE solution. In Fig. \ref{Mode Shapes}, we compare the displacements of an arbitrarily selected frequency-phase velocity pair $(f,c_{ph})\approx(2.02, 6.18)$, taken from Fig. \ref{Dispersion Curves}, again showing a good match. The SCM solutions are obtained as an eigenvector of a polynomial eigenvalue problem, and any scalar multiple remains a solution. To fix upon a solution we impose $u_x$ at the interface $y=-d$ to be the same for both SCM and DISPERSE solutions. 

\section{Conclusion}\label{Conclusion}
The leaky Lamb wave problem, for a plate loaded on both sides by two different fluids, 
 is an exemplar of leaky waves and exhibits the key difficulty of studying such systems; the nonlinear eigenvalue problem is difficult to solve, a neat change of variables to transform it is not easily available, the exponential growth poses a numerical challenge to the retrieval of accurate solutions. To overcome this inherent difficulty a mapping of the fluid domains onto paths in complex domains was performed; the exponential growth is then built into this and we can reformulate  
as a polynomial eigenvalue problem solved using standard solvers; thence we obtain the complete physical solution in a robust and accurate manner. 
The spectral collocation methods, based on Chebyshev approaches to discretise both the open and closed domains of our system, provides  our method with speed, accuracy and robustness whilst keeping the eigenvalue systems compact. The formulation leads to systems amenable to standard eigensolvers and we are able to obtain all the complex wavenumbers and displacement fields of leaky Lamb waves for any given frequency. \\ \\ \indent The complex mapping and the treatment of the exponential growth of the leaky waves of our system is not unique to the planar elastic waveguiding problem: the methodology we have introduced can be extended to multilayered elastic systems of plain or cylindrical geometries. 
The underlying SCM technique is also general and can be extended to more complicated geometries and guiding structures, and also to more general media i.e. anisotropic or viscoelastic guides, and thereby together with the leaky wave SCM provides a methodology for all wave systems of this generic type including those in electromagnetism.

\begin{acknowledgments}
EG and RVC are supported by the EU H2020 FET
Open ‘‘Boheme” Grant No. 863179
\end{acknowledgments}


\bibliography{Bibliography}

\end{document}


\setstretch{0.7}
\title[]{[Supplementary Material] Leaky wave characterisation using spectral methods}


\author{Evripides Georgiades}
\email{evripides.georgiades18@imperial.ac.uk}
\affiliation{Department of Mechanical Engineering, Imperial College London, London, SW7 1AY, United Kingdom}

\author{Michael J.S. Lowe}
\affiliation{Department of Mechanical Engineering, Imperial College London, London, SW7 1AY, United Kingdom}

\author{Richard V. Craster}
\altaffiliation{Also at: Department of Mathematics, Imperial College London, London, SW7 1AY, United Kingdom}
\affiliation{Department of Mechanical Engineering, Imperial College London, London, SW7 1AY, United Kingdom}





\date{\today} 


\maketitle
\nolinenumbers
\section*{Overview of Supplementary Material}
In this supplementary document, we give the derivations of the momentum equations and interface conditions used in the main article, a brief overview of the theory of Differentiation Matrices as well as a discussion on the choice of the complex parameters $\zeta_{f_j}$.

\section*{Momentum Equations}
From the main article,  under the assumption of plane harmonic waves in the direction of propagation, $x$, and in time, $t$, the displacements in the solids are given by 
\begin{align}
\label{supplementary u_x} \bar{u}_x(x,y,t)&=u_x(y)e^{i(k_xx-\omega t)}, \\
\label{supplementary u_y} \bar{u}_y(x,y,t)&=u_y(y)e^{i(k_xx-\omega t)},
\end{align}

\noindent and the displacements in the fluids are given by

\begin{equation} \label{supplementary displacement vectors fluids} 
  \boldsymbol {u_{f_1}}=\begin{pmatrix}
\bar{u}_{x_{f_1}} \\
\bar{u}_{y_{f_1}} \\
0 \end{pmatrix}=\nabla \bar{\phi}_{f_1}, 
  \hspace{0.5cm} \boldsymbol {u_{f_2}}=\begin{pmatrix}
\bar{u}_{x_{f_2}} \\
\bar{u}_{y_{f_2}} \\
0 \end{pmatrix}=\nabla \bar{\phi}_{f_2},
\end{equation}

\noindent with 

\begin{align}
\label{supplementary phi_f1} \bar{\phi}_{f_1}(x,y,t)&=\phi_{f_1}(y)e^{i(k_xx-\omega t)}, \\
\label{supplementary phi_f2} \bar{\phi}_{f_2}(x,y,t)&=\phi_{f_2}(y)e^{i(k_xx-\omega t)},
\end{align} 

Substitution of Eqs. \eqref{supplementary phi_f1} and  \eqref{supplementary phi_f2} into Eq. \eqref{supplementary displacement vectors fluids} gives the form of the displacements in the fluids, in terms of the fluid wave potentials, as 

\begin{align} 
   \label{supplementary ufx} \bar{u}_{x_{f_j}}&= ik_x\phi_{f_j}(y)e^{i(k_xx-\omega t)}\\
\label{supplementary ufy} \bar{u}_{y_{f_j}}&=\frac{\partial \phi_{f_j}}{\partial y}e^{i(k_xx-\omega t)}.
\end{align}

For a description of wave motion in the immersed waveguide, we consider the momentum equations\citep{Auld1973AcousticSolids,Nayfeh1988PropagationComposite}

\begin{align} \label{supplementary momentum equation ux}
    \frac{\partial\sigma_x}{\partial x}+\frac{\partial\sigma_{xy}}{\partial y}=\rho\frac{\partial^2 \bar{u}_x}{\partial t^2}, \\
    \label{supplementary momentum equation uy} \frac{\partial\sigma_y}{\partial y}+\frac{\partial\sigma_{xy}}{\partial x}=\rho\frac{\partial^2 \bar{u}_y}{\partial t^2},
\end{align}

\noindent where the stresses are given by

\begin{align}
    \label{supplementary stress x} \sigma_x&=(\lambda+2\mu)\frac{\partial \bar{u}_x}{\partial x}+\lambda \frac{\partial \bar{u}_y}{\partial y}, \\
    \label{supplementary stress y} \sigma_y&=(\lambda+2\mu)\frac{\partial \bar{u}_y}{\partial y}+\lambda \frac{\partial \bar{u}_x}{\partial x}, \\
    \label{supplementary stress xy} \sigma_{xy}&=\mu\left(\frac{\partial \bar{u}_y}{\partial x}+\frac{\partial \bar{u}_x}{\partial y}\right).
\end{align}

\noindent A substitution of Eqs. \eqref{supplementary u_x} and \eqref{supplementary u_y} in Eqs. \eqref{supplementary stress x}-\eqref{supplementary stress xy} gives the stresses in the solid, in terms of the $y-$dependent component of the solid displacements, as 

\begin{align}
\label{supplementary stress x solid}
    \sigma_x&=(\lambda+2\mu)ik_xu_x+\lambda\frac{\partial u_y}{\partial y}, \\
\label{supplementary stress y solid}
    \sigma_y&=(\lambda+2\mu)\frac{\partial u_y}{\partial y}+\lambda ik_xu_x, \\
\label{supplementary stress xy solid}
    \sigma_{xy}&=\mu\left(ik_xu_y+\frac{\partial u_x}{\partial y}\right).
\end{align}

Then, a substitution of Eqs. \eqref{supplementary u_x} and \eqref{supplementary u_y} and \eqref{supplementary stress x solid}-\eqref{supplementary stress xy solid} into the momentum equations Eqs. \eqref{supplementary momentum equation ux} and \eqref{supplementary momentum equation uy}, results in the momentum equations for the solid, in terms of the $y-$dependent component of displacements, used in the main article. 

Due to the modeling of inviscid fluids as isotropic solids, the derivation of the momentum equations for wave motion in the inviscid fluids is similar to the one presented for the solid waveguide. We start by considering the momentum equations \citep{Auld1973AcousticSolids,Nayfeh1988PropagationComposite}

\begin{align} \label{supplementary momentum equation ufx}
    \frac{\partial\sigma_{x_{f_j}}}{\partial x}+\frac{\partial\sigma_{xy_{f_j}}}{\partial y}=\rho_{f_j}\frac{\partial^2 \bar{u}_{x_{f_j}}}{\partial t^2}, \\
    \label{supplementary momentum equation ufy} \frac{\partial\sigma_{y_{f_j}}}{\partial y}+\frac{\partial\sigma_{xy_{f_j}}}{\partial x}=\rho_{f_j}\frac{\partial^2 \bar{u}_{y_{f_j}}}{\partial t^2},
\end{align}

\noindent for $j=1,2$ corresponding to each fluid. The stresses in the fluids are given by

\begin{align}
    \label{supplementary stress fx & fy} \sigma_{x_{f_j}}=\sigma_{y_{f_j}}&=\lambda_{f_j}\left(\frac{\partial \bar{u}_{x_{f_j}}}{\partial x}+\frac{\partial \bar{u}_{y_{f_j}}}{\partial y}\right), \\
    \label{supplementary stress fxy} \sigma_{xy_{f_j}}&=0.
\end{align}

\noindent The assumption of volumetric stress in the two isotropic fluids \citep{Auld1973AcousticSolids}, allows us to write the  stresses in the fluids, in terms of the $y-$dependent components of the fluid wave potentials, as

\begin{align}
    \label{supplementary stress x & y fluid} \sigma_{x_{f_j}}=\sigma_{y_{f_j}}&=-\rho_{f_j}\omega^2\phi_{f_j}, \\
    \label{supplementary stress xy fluid} \sigma_{xy_{f_j}}&=0.
\end{align}

With a final substitution of Eqs. \eqref{supplementary ufx}, \eqref{supplementary ufy}, \eqref{supplementary stress x & y fluid} and \eqref{supplementary stress xy fluid} in Eqs. \eqref{supplementary momentum equation ufx} and \eqref{supplementary momentum equation ufy} results in the same equation for both. This is the momentum equation for the fluids, in terms of the $y-$dependent component of wave potentials, used in the main article.

\section*{Interface Conditions}
For a two-sided fluid-loaded plate the interface conditions were the continuation of traction and of normal displacements at the two fluid-solid interfaces \citep{Rose2014UltrasonicMedia}. Those translate to 

\begin{align}
    \label{supplementary ic 1}(\sigma_{y_{f_j}}-\sigma_y)|_{y=y_j}=0 \\
    \label{supplementary ic 2}\sigma_{xy}|_{y=y_j}=0 \\
    \label{supplementary ic 3}(u_{y_{f_j}}-u_y)|_{y=y_j}=0
\end{align}

\noindent with $y_1=-d$, $y_2=d$, and $u_{y_{f_j}}$ the $y-$dependent component of $\bar{u}_{y_{f_j}}$. From Eq. \eqref{supplementary ufy}, we have that \begin{equation}
    u_{y_{f_j}}=\frac{\partial \phi_{f_j}}{\partial y}
\end{equation}.

\noindent Then, a direct substitution of Eqs. \eqref{supplementary stress y},\eqref{supplementary stress xy},\eqref{supplementary stress x & y fluid} and \eqref{supplementary ufy} in Eqs. \eqref{supplementary ic 1}-\eqref{supplementary ic 3} yields the interface conditions of the main article, in terms of the $y-$dependent components of wave displacements and wave potentials. 

\section*{Differentiation Matrices} \label{DM}
 Consider a continuous function $g(y)$, with $y$ in some domain $\mathcal{C}$, whose derivative is to be approximated. With spectral collocation we choose weighted interpolants, \citep{Fornberg1996AMethods,Boyd2000ChebyshevMethods}

\begin{equation}
    \label{interpolants} g(y) \approx p_{N-1}(y)=\sum\limits_{i=1}^N \frac{\alpha(y)}{\alpha(y_i)}\psi_i(y)g_i,
\end{equation}

\noindent where $\{y_i\}_{i=1}^N$ is a set of $N$ distinct points in $\mathcal{C}$, $\alpha(x)$ is a weight function, $g_i=g(x_i)$ and $\{\psi_i(y)\}_{i=1}^N$ are a set of interpolating functions that satisfy $\psi_i(y_k)=\delta_{ik}$ (Kronecker delta); due to the choice of interpolating functions and weights, we have that $g(y_j)=p_{N-1}(y_j)$ for $i=1, \dots, N.$ This implies a zero residual error at the $N$ distinct points $\{y_j\}_{j=1}^N$, known as the collocation points. Furthermore, the interpolants $\psi_j(y)$ can be chosen to be global functions, as opposed to piece-wise polynomials used in FE models, leading to much higher accuracy, with a convergence rate of up to $O(e^{-cN})$ for smooth solutions \citep{Weideman2000ASuite}; this is known as spectral accuracy.

With this choice for $g(y)$, we obtain the $l^{th}$ derivative of $g$ at the $N$ collocation points as

\begin{equation} \label{approximate diferentiation}
    g^{(l)}(y_k) \approx \sum \limits_{i=1}^N\frac{d^l}{dy^l} \left [ \frac{\alpha(y)}{\alpha(y_i)}\psi_i(y)\right]_{y=y_k}g_i,
\end{equation}

\noindent for $k=1,\dots, N$. Eq. \eqref{approximate diferentiation} is summarised in a single matrix, known as the Differentiation Matrix in $\mathcal{C}$, given by 

\begin{equation} \label{differentiation matrix definition}
    D^{(l)}_{k,i}=\frac{d^l}{dy^l} \left [ \frac{\alpha(y)}{\alpha(y_i)}\psi_i(y)\right]_{y=y_k}.
\end{equation}

As in the main article, when $\boldsymbol{g}$ is the vector of $f$ evaluated at the $N$ collocation points and $\boldsymbol{g}^{(l)}$ is the vector of their $l^{th}$ derivatives, we can approximate $\boldsymbol{g}^{(l)}$ by

\begin{equation}
    \boldsymbol{g}^{(l)} \approx D^{(l)}\boldsymbol{g}.
\end{equation}
\section*{Choice of complex $\zeta_{f_j}$}
Under the maps $h_{f_1}$ and $h_{f_2}$, the $y-$dependent components of wave potentials of the complex valued problem are given by the equations 

\begin{align}
    \label{mapped fluid wave potentials 1}
    &\phi_{f_1}(y)=A_1\exp({-ik_{y_1}\zeta_{f_1}y}), \\
    \label{mapped fluid wave potentials 2}
    &\phi_{f_2}(y)=A_2\exp({ik_{y_2}\zeta_{f_2}y}).
\end{align}

\noindent where $A_1, A_2$ are constants and $k_{y_j}\in\mathbb{C}$ lie in the $4^{th}$ complex quadrant. Furthermore, in Eq. \eqref{mapped fluid wave potentials 1}, $y$ is in the real half line $[-\infty,-d]$ and in Eq. \eqref{mapped fluid wave potentials 2}, $y$ is in $[d,\infty].$ 

For our method to converge, we need to make a choice of $\zeta_{f_j}\in\mathbb{C}$ such that the solutions of Eqs. \eqref{mapped fluid wave potentials 1} and \eqref{mapped fluid wave potentials 2} decay exponentially. We demonstrate the choice for $\zeta_{f_1}$, with the choice of $\zeta_{f_2}$ following a similar argument. 

We start by writing $k_{y_1}=\alpha+i\beta$ and $\zeta_{f_1}=\gamma+i\delta$, with $\alpha,\gamma,\delta \in\mathbb{R}_{>0}$ and $\beta\in\mathbb{R}_{<0}$. With this notation, equation \eqref{mapped fluid wave potentials 1} reads 
\begin{equation}\label{potential for choice of zeta 1}
\begin{split}
     \phi_{f_1}(y)&=A_1\exp(-ik_{y_1}\zeta_{f_1}y)\\
    &=A_1\exp[-i(\alpha+i\beta)(\gamma+i\delta)y] \\
    &=A_1\exp[-iy(\alpha \gamma+i\alpha \delta+i\beta \gamma-\beta \delta)] \\
    &=A_1\exp[-iy(\alpha \gamma-\beta \delta)]\exp[y(\alpha \delta+\beta \gamma)].
\end{split}
\end{equation}

It follows that for a decaying potential in Eq. \eqref{potential for choice of zeta 1}, we must have that 

\begin{equation}
    \begin{split}
        y(\alpha \delta+\beta \gamma)&<0 \\
        \alpha \delta+\beta \gamma&>0 \\
        \delta&>-\gamma(\beta/\alpha).
    \end{split}
\end{equation}

\linenumbers
\nolinenumbers

\bibliography{Bibliography}